# Quantum Melting of a Wigner crystal of Rotating Dipolar Fermions in the Lowest Landau Level

Szu-Cheng Cheng[*]

Department of Physics, Chinese Culture University, Taipei, Taiwan, Republic of China

Shih-Da Jheng and T. F. Jiang

Institute of Physics, National Chiao Tung University, Hsinchu, Taiwan 30010

Abstract

We have investigated the behavior and stability of a Wigner crystal of rotating dipolar fermions in two dimensions. Using an ansatz wave function for the ground state of rotating two-dimensional dipolar fermions, which occupy only partially the lowest Landau level, we study the correlation energy, elastic moduli and collective modes of Wigner crystals in the lowest Landau level. We then calculate the mean square of the displacement vector of Wigner crystals. The critical filling factor, below which the crystalline state is expected, is evaluated at absolute zero by use of the Lindeman's criterion. We find that the particle (hole) crystal is locally stable for $\nu <1/15$ ($14/15<\nu<1$), where the stable regime of the crystal is much narrower than the result from Baranov, Fehrmann and Lewenstein, [Phys. Rev. Lett. **100**, 200402 (2008)].

---

[*] Electronic address: sccheng@faculty.pccu.edu.tw; Fax: +886-2-28610577

# I. INTRODUCTION

There is a remarkable progress in the experiment to manipulate quantum many-body states in the ultra cold atomic gases. Systems with well understood and controllable interactions are available. In most researches so far, the major systems that have been studied are described by a contact potential that is characterized by the *s*-wave scattering [1-3]. The successful creation of chromium Bose-Einstein condensate (BEC) [4, 5] and progress in realization heteronuclear polar molecules [6-11] has been the subject of growing interest in quantum degenerate dipolar gases. The anisotropic character and long-rang nature of dipole-dipole interactions makes the dipolar systems different from the systems described by contact interactions [12, 13]. For the BEC, new quantum phases are predicted [14, 15]. There is a quantum phase transition from a gas to a solid phase when the density increases. The influence of the trapping geometry on the stability of the BEC and the effect of the dipole-dipole interactions on the excitation spectrum are investigated [16]. Vortex lattices of rotating dipolar BECs exhibit novel bubble, stripe, and square structures [17]. For Fermi gases, the *s*-wave scattering is inhibited due to the Pauli Exclusion Principle. This principle also provides a strong suppression of three-body losses [18], and consequently the strong interactions of fermions are achieved at the Feshbach resonance where the scattering length takes a divergent value [19, 20]. Bond pairs of fermions with resonant interaction are formed and the system of a Fermi gas is transformed into a bosonic gas of molecules [21, 22]. The observed pairing of fermions provides the crossover between the Bardeen-Cooper-Schreiffer regime of weakly-paired, strongly overlapping Cooper pairs, and the BEC regime of tightly bound, weakly-interacting diatomic molecules [23]. It is also possible to explore the potentially strong correlations of fermions induced by the dipolar interaction [24]. There are dipolar-induced superfluidity [25, 26] and strongly correlated states in rotating dipolar Fermi gases [13, 27].

In a system whose potential energy from interactions dominates the kinetic energy, the system will adopt a configuration of the lowest potential energy and crystallize into a lattice that is termed the Wigner crystal (WC) [28]. Rotating fermions with the inertial mass $M$ and the rotational frequency $\omega$ feel the Coriolis force in the rotating frame. The Coriolis force on rotating gases is identical to the Lorentz force of a charged particle in a magnetic field. Quantum-mechanically, energy levels of a charged particle in a uniform magnetic field are named as Landau levels and display discrete energy levels of a harmonic oscillator [29]. There is a large degeneracy in the Landau level. It is convenient to define the filling factor ν, $\nu \equiv 2\pi\rho l_0^2$, to measure the fraction of Landau levels being occupied by particles, where $\rho$ and $l_0 = \sqrt{\hbar/2M\omega}$ stand for the average density and the magnetic length, respectively. The discreteness of the eigenvalue spectrum is essential for the occurrence of the integral quantum Hall effect [30]. In the lowest Landau level, the kinetic energy of rotating dipolar gases is frozen and the dipolar interactions create strong correlations on particles. Therefore, the potential energy from dipolar interactions dominates the kinetic energy in the lowest Landau level and, except at certain filling factors, we shall expect that rotating dipolar gases will crystallize into a WC. Baranov, Fehrmann and Lewenstein found that it is feasible of forming the crystal phase for $\nu < \frac{1}{7}$ [27]. Their study is valid in the regime for $\nu < \frac{1}{2}$. Physical properties of rotating dipolar gases in the interval between $\frac{1}{2} < \nu < 1$ is still lack of investigation. It was shown that a hole crystal for a two-dimensional (2D) electron gas could exist in the interval between $\frac{1}{2} < \nu < 1$ [31]. The hole-crystal state is described by the completely filled Landau level plus a hole crystal with the hole density $(1-\nu)$. It is interesting to study physical properties and their stability from both kinds of crystals, i.e., electron and hole crystals in the lowest Landau level.

In this paper we construct an ansatz wave function of the WC of rotating 2D dipolar fermions, which occupy only partially the lowest Landau level. Within the Hartree-Fock approximation and ignoring the Landau-level mixing effect, we study the correlation energy and elastic properties of the WC. We show that the correlation energy of a particle crystal is lower and higher than the correlation energy of a hole crystal for $\nu < \frac{1}{2}$ and $\nu > \frac{1}{2}$, respectively. We also examine the stability of a Wigner crystal by applying the Lindeman's criterion that the crystal will melt if the mean square of the displacement vector exceeds some critical value. The paper is organized as follows. In Section II we construct an ansatz of the WC. The overlapping integral of trial wave functions between two different sites of the lattice is non-zero. We then develop a method of handling trial non-orthogonal wave functions by the Löwdin formula. From this formula we can calculate the correlation energy of a WC accurately. Calculation results of the correlation energy are given in Section III. In Section IV we find the compression and shear moduli of the WC in the lowest Landau level. We then calculate collective modes of the WC and show their long-wavelength dispersions. In Section V we derive the mean square of the displacement vector in the harmonic approximation. From the derivation, we can find the stable region of the WC. Finally, Section VI contains a summary of our results along with our conclusions.

## II. HARTREE-FOCK APPROXIMATION AND LÖWDIN FORMULA

We consider polarized dipolar fermions in a rotating cylindrical trap and polarized along the negative $\mathbf{e}_z$ direction of rotation, where $\mathbf{e}_z$ is the unit vector of the $z$-axis. The two-body dipolar interaction potential is

$$\Delta(\mathbf{\eta}) = D \frac{1 - 3z^2/\mathbf{\eta}^2}{|\mathbf{\eta}|^3}, \tag{1}$$

where $D$ is a measure of the strength of the dipolar interactions, and $\mathbf{\eta} = (x, y, z)$ is a position vector of a particle. For the sake of simplicity, we will consider a pancake system such that the motion of rotating fermionic gases along the $z$-axis is frozen to the ground state of the axial harmonic oscillator whose ground-state wave function is

$$\phi(z) = \exp(-z^2/2d^2)/(\pi d^2)^{1/4}, \tag{2}$$

where $d = \sqrt{\hbar/M\omega_z}$ is the extension of dipolar gases in the axial direction and $\omega_z$ is the trapping frequency along the z-axis. The quasi-2D dipole-dipole interaction between fermions is given by

$$V(\mathbf{r}_1 - \mathbf{r}_2) = \int dz_1 dz_2 |\phi(z_1)|^2 |\phi(z_2)|^2 \Delta(\boldsymbol{\eta}_1 - \boldsymbol{\eta}_2), \tag{3}$$

where $\mathbf{r} = (x, y)$.

The formal development of the Hartree-Fock ground-state energy for a 2D WC induced by a strong magnetic field has been presented elsewhere [32]. We shall apply previous formalisms and take the cyclotron energy $\hbar\omega_c = 1$ and the magnetic length $l_0 = 1$, where the cyclotron frequency $\omega_c = 2\omega$. The Hamiltonian of a quasi-2D rotating dipolar Fermi gas is

$$H = \frac{1}{2}\sum_{i=1}^{N}\left[-i\nabla_i - \frac{1}{2}\mathbf{r}_i \times \mathbf{e}_z\right]^2 + \frac{\Omega}{2L^2}\sum_{i\neq j=1}^{N} V(\mathbf{r}_i - \mathbf{r}_j), \tag{4}$$

and

$$V(\mathbf{r}_i - \mathbf{r}_j) = \sum_{\mathbf{q}} V(\mathbf{q}) e^{i\mathbf{q}\bullet(\mathbf{r}_i - \mathbf{r}_j)}, \tag{5}$$

where $N$ is the total number of particles, $\mathbf{q}$ is the wave vector, $L$ is the length of the system and $\Omega \equiv (D/l_0^3)/\hbar\omega_c$. The $V(\mathbf{q})$ in Eq. (5) is the Fourier transform of $V(\mathbf{R})$ and

$$V(\mathbf{q}) = \frac{4\sqrt{2\pi}}{3(d/\ell)} - 2\pi q \exp(\xi^2)\, \text{Efrc}(\xi), \tag{6}$$

where $\xi = qd/\sqrt{2}l_0$ and $\text{Erfc}(\xi) = 1 - \frac{2}{\sqrt{\pi}}\int_0^\xi \exp(-t^2)dt$. Eq. (6) was used in the study of the stability of 2D BEC with dominant dipole-dipole interactions [16]. Note that the contact interaction term in $V(\mathbf{q})$ can be ignored due to the Pauli Exclusion Principle of fermions.

The lowest Landau level wave function localized around $\mathbf{r} = \mathbf{R}$ is given by

$$\psi_{\mathbf{R}}(\mathbf{r}) = \frac{1}{\sqrt{2\pi}} \exp\left[-\frac{1}{4}(\mathbf{r}-\mathbf{R})^2 + \frac{1}{2}i(\mathbf{r}\times\mathbf{R})\cdot\mathbf{e}_z\right], \tag{7}$$

which is the eigenstate, with eigenvalue $\frac{1}{2}\hbar\omega_c$, of the kinetic energy operator. This wave function was proposed by Maki and Zotos to study the correlation, the shear moduli, and collective excitation modes of the WC for electrons partially occupied the lowest Landau level [31]. Our ansatz wave function for the ground state of the WC is a Slater determinant constructed by the wave function of Eq. (7) located at the regular 2D lattice points $\mathbf{R}_j$. We have

$$\Psi(\{\mathbf{r}_i\}) = \frac{1}{\sqrt{N!}} \det\left|\psi_{\mathbf{R}_j}(\mathbf{r}_i)\right|. \tag{8}$$

From the previous study [33] we know that a 2D lattice with the triangular geometry structure has the lowest classical ground-state energy. Therefore, we choose the lattice site $\mathbf{R}_j$ of the WC as

$$\mathbf{R}_j = j_1 \mathbf{a}_1 + j_2 \mathbf{a}_2, \tag{9}$$

where $\mathbf{a}_1$ and $\mathbf{a}_2$ are the primitive translation vectors of the lattice; and $j_1$ and $j_2$, are any integers to which we refer collectively as $j$. We also define a lattice reciprocal to the direct lattice defined by Eq. (9) as the set of points given by the vectors

$$\mathbf{G}_m = m_1 \mathbf{b}_1 + m_2 \mathbf{b}_2, \tag{10}$$

where $\mathbf{b}_1$ and $\mathbf{b}_2$ are the primitive translation vectors of the reciprocal lattice; and $m_1$ and $m_2$, are any integers to which we refer collectively as $m$. For a triangular lattice

$$\mathbf{a}_1 = a(1,\ 0), \quad \mathbf{a}_2 = a\left(\frac{1}{2},\ \frac{\sqrt{3}}{2}\right),$$

$$\mathbf{b}_1 = \frac{2\pi}{a}\left(1, -\frac{\sqrt{3}}{3}\right), \quad \mathbf{b}_2 = \frac{2\pi}{a}\left(0, \frac{2\sqrt{3}}{3}\right), \tag{11}$$

where $a$ is the lattice constant. The area $A_c$ of the primitive unit cell of the triangular lattice is $\sqrt{3}a^2/2$. Since the particle density $\rho$ is related to the filling factor $\nu$, the lattice constant $a$ is given by $a = \sqrt{(4\pi l_0^2)/(\sqrt{3}\nu)}$.

The overlapping integral of trial wave functions between two different sites of the lattice is non-zero. Our wave functions are normalized but not orthogonal each other for different sites, we cannot use the Hartree-Fock equation for orthonormal wave functions to calculate the ground-state energy. We have to consider this non-orthogonal situation seriously. Let us define a $\Theta$ matrix with elements [34]

$$\theta_{ij} = \langle i | j \rangle = \int d\mathbf{r}\, \psi^*_{\mathbf{R}_i}(\mathbf{r}) \psi_{\mathbf{R}_j}(\mathbf{r}) . \tag{12}$$

The total energy per particle $E$ of the WC can be expressed in terms of elements $T_{ij}$ of matrix $\mathbf{T}$ and $\mathbf{T} = \Theta^{-1}$ is the reciprocal matrix of $\Theta$:

$$E = K + V_d - V_{ex}, \tag{13}$$

where $K$ is the kinetic energy per particle, $V_d$ is the direct interaction energy per particle, and $V_{ex}$ is the exchange energy per particle, respectively; and they are given as

$$K = \frac{1}{N}\sum_{ij} T_{ji} \left\langle i \left| \frac{1}{2}\left[-i\nabla - \frac{1}{2}\mathbf{r}\times\mathbf{e}_z\right]^2 \right| j \right\rangle, \tag{14}$$

$$V_d = \frac{\Omega}{2NL^2}\sum_{ijnk} T_{ji} T_{kn} \langle i, n | V(\mathbf{r}_1 - \mathbf{r}_2) | j, k \rangle, \tag{15}$$

and

$$V_{ex} = \frac{\Omega}{2NL^2} \sum_{ijnk} T_{jn}T_{ki} \langle i,n|V(\mathbf{r}_1 - \mathbf{r}_2)|j,k\rangle. \tag{16}$$

The formula $V_d - V_{ex}$ is essentially Löwdin's formula [34] for the non-orthogonal wave functions. If wave functions are orthonormal, then $T_{ij} = \delta_{ij}$. The formula $V_d - V_{ex}$ is just a well-known formula in the Hartree-Slater-Fock theory. From the translational invariant properties of Löwdin's formula [32], we have

$$T_{ij} = T_{(i-j)0} \exp\left(\frac{i}{2}\mathbf{R}_i \times \mathbf{R}_j \bullet \mathbf{e}_z\right). \tag{17}$$

Using Eq. (17) and with aid of the transformation

$$\frac{1}{L^2}\sum_j e^{i\mathbf{q}\bullet\mathbf{R}_j} = \frac{1}{A_c}\sum_{\mathbf{G}} \delta_{\mathbf{G},\mathbf{q}}, \tag{18}$$

where the $\mathbf{G}$'s are the reciprocal lattice vectors, equations (14), (15) and (16) can be rewritten as

$$K = \frac{1}{2}\sum_j \langle 0|j\rangle T_{j0} = \frac{1}{2}, \tag{19}$$

$$V_d = \frac{\Omega}{2L^2}\sum_{jk} T_{j0}T_{k0}\langle 0,0|\sum_i V(\mathbf{r}_1 - \mathbf{r}_2 - \mathbf{R}_i)|j,k\rangle$$

$$= \frac{\Omega}{2A_c}\sum_{jk,\mathbf{G}} V(\mathbf{G})T_{j0}T_{k0}\langle 0,0|e^{i\mathbf{G}\bullet(\mathbf{r}_1-\mathbf{r}_2)}|j,k\rangle,$$

$$\tag{20}$$

and

$$V_{ex} = \frac{\Omega}{2L^2}\sum_{ijk} T_{j0}T_{k0}\langle 0,i|V(\mathbf{r}_1 - \mathbf{r}_2)|j+i,k\rangle e^{\frac{-i}{2}(\mathbf{R}_i \times \mathbf{R}_j)\bullet\mathbf{e}_z}, \tag{21}$$

where $|0\rangle$ and $|j+i\rangle$ denote particle states at $\mathbf{R}=0$ and $\mathbf{R} = \mathbf{R}_j + \mathbf{R}_i$, respectively.

# III. CORRELATION ENERGY OF THE WIGNER CRYSTAL

We chose 61 lattice sites around the origin to calculate the **T** matrix numerically. Using the formula $V_d - V_{ex}$ and ignoring the constant kinetic energy, we can calculate the correlation energy per particle $V_c(\nu) = V_d - V_{ex}$ as a function of the filling factor $\nu$ of the WC. To test our calculations we can do two things: Firstly, we compare our calculations with the WC energy $V_B$ calculated by Baranov *et al.* [27]. In Fig. 1, we show our results of the correlation energy for the dipolar gas with thickness $d/l_0 = 0$. Our calculations are consistent with the results from Baranov *et al.* in the small filling factor regime. As the filling factor is increasing, our calculations start deviating from the results from Baranov *et al.*, which is valid in the small filling-factor regime. Secondly, as the lowest Landau level completely filled, i.e., $\nu=1$, the correlation energies of the dipolar gas and WC should be equal. We got $V_{gas}(\nu=1) = 0.62665\ D/l_0^3$ and $V_c(\nu=1) = 0.62663\ D/l_0^3$ as $d/l_0 = 0$. From the above comparisons we can see that Löwdin's formula can give us accurate correlation energies of the WC in the lowest Landau level.

In Fig. 1, we also exhibit the correlation energy of the hole crystal $V_{hole}(\nu)$ as a function of the filling factor $\nu$. A hole-crystal state is described by the completely filled Landau level plus a hole crystal with the hole density $(1-\nu)$. The correlation energy of the hole crystal $V_{hole}(\nu)$ is given by [31]

$$V_{hole}(\nu) = V_c(1-\nu) + (2\nu - 1)\ V_{gas}(\nu = 1), \tag{22}$$

where the second term comes from the exchange energy between holes and the underlying dipolar gas. A hole crystal has a lower correlation energy than a particle crystal for $\nu > \frac{1}{2}$. Furthermore the separate correlation-energy slopes of particle and hole crystals at $\nu=\frac{1}{2}$ are different. Therefore, the phase transition of particle and hole crystals at $\nu=\frac{1}{2}$ is the first order.

The correlation energies of particle and hole crystals for $d/l_0 = 0.3$ and $0.7$ are shown in Fig. 2. The correlation energy of dipolar gases with finite thickness has the same qualitative behavior as the thickness $d/l_0 = 0$. The correlation energy is increasing as the filling factor is increasing. The hole crystal has the lowest correlation energy for $\nu > \frac{1}{2}$. The correlation energy is decreasing as the thickness of dipolar gases is increasing. This behavior implies that the WC with a larger thickness is less stable than the WC with a smaller thickness.

## IV. DYNAMICAL MATRIX AND COLLECTIVE MODES OF THE WIGNER CRYSTAL

Having obtained the correlation energy of the WC, we now would like to study the elastic properties and collective modes of the WC. In order to do so, we still have to derive an effective interaction between the particles, which then will allow us to find the dynamical matrix, the compression and shear moduli of the WC.

Going back to equations (20) and (21), if we furthermore define elements

$$\tau(i,j,\mathbf{q},\mathbf{R}) = e^{-\frac{1}{4}\left(\mathbf{R}_i^2+\mathbf{R}_j^2\right)-\mathbf{q}^2} \times e^{\frac{1}{2}\mathbf{q}\bullet\left(\mathbf{R}_j-\mathbf{R}_i\right)\times\mathbf{e}_z} \times e^{\frac{i}{2}\mathbf{q}\bullet\left(\mathbf{R}_i-\mathbf{R}_j\right)}$$

$$\times e^{-\frac{1}{2}\mathbf{R}^2+\frac{1}{2}\mathbf{R}\bullet(\mathbf{R}_i-\mathbf{R}_j)} \times e^{\mathbf{q}\bullet\mathbf{R}\times\mathbf{e}_z+\frac{i}{2}\mathbf{R}\times(\mathbf{R}_i-\mathbf{R}_j)\bullet\mathbf{e}_z}, \tag{23}$$

and

$$\tau(\mathbf{q}) = \sum_{ij} T_{i0}T_{j0}\ \tau(i,j,\mathbf{q},0), \tag{24}$$

then we can rewrite the correlation energy in the form:

$$V = \frac{1}{2} \sum_{m \neq n} U(\mathbf{R}_m - \mathbf{R}_n), \quad (25)$$

where we introduced the effective interaction potential $U(\mathbf{R})$ between particles localized at lattice sites, which in real space is given by

$$U(\mathbf{R}) = \frac{1}{L^2} \sum_{\mathbf{q}} V(\mathbf{q}) \left[ \tau(\mathbf{q}) e^{i\mathbf{q}\cdot\mathbf{R}} - \sum_{ij} T_{i0} T_{j0} \tau(i,j,\mathbf{q},\mathbf{R}) \right]. \quad (26)$$

We shall use the above expression of the interaction potential to study the elastic properties of the WC.

The derivation of the elastic moduli associated with the effective interaction potential in Eq. (25) proceeds as follows. Firstly, we evaluate the dynamical matrix $\Phi_{\alpha\beta}(\mathbf{k})$, $\alpha, \beta = x, y$, which is given by

$$\Phi_{\alpha\beta}(\mathbf{k}) = \sum_{\mathbf{R} \neq 0} e^{-i\mathbf{k}\cdot\mathbf{R}} \Phi_{\alpha\beta}(\mathbf{R}), \quad (27)$$

where

$$\Phi_{\alpha\beta}(\mathbf{R}) = -\frac{\partial^2 U(\mathbf{R})}{\partial R^\alpha \partial R^\beta}. \quad (28)$$

Expanding $\Phi_{\alpha\beta}(\mathbf{k})$ in $\mathbf{k}$ around $\mathbf{k}=0$ leads to the form

$$\Phi_{\alpha\beta}(\mathbf{k}) = (C_{11} - C_{66}) k_\alpha k_\beta + C_{66} k^2 \delta_{\alpha\beta}, \quad (29)$$

where $C_{11}$ and $C_{66}$ are the compression and shear moduli of the WC, respectively. $C_{11}$ and $C_{66}$ can be extracted from the expression of $\Phi_{\alpha\beta}(\mathbf{k})$ according to

$$C_{11} = \frac{1}{2} \frac{d^2}{dk_x^2} \Phi_{xx}(k_x, k_y = 0), \quad (30)$$

and

$$C_{66} = \frac{1}{2} \frac{d^2}{dk_y^2} \Phi_{xx}\left(k_x = 0, k_y\right). \tag{31}$$

The constants $C_{11}$ and $C_{66}$ are evaluated numerically. Figures 3 and 4 show the variation of the shear and compression moduli of particle crystals as a function of the sample thickness $d/l_0$ in the filling factors $v = \frac{1}{7}, \frac{1}{5}$, and $\frac{1}{3}$. $C_{66}$ is a decreasing function of $d/l_0$, implying that the WC with a larger thickness is less stable than the WC with a smaller thickness. It is consistent with the result from the correlation energy that the stability of the WC is decreasing as $d/l_0$ is increasing. For the compression moduli, $C_{11}$ becomes smaller as $d/l_0$ is increasing.

The compression moduli of the WC as a function of $v$ in the sample thicknesses $d/l_0 = 0$, 0.3, and 0.7 are shown in Fig. 5. For smaller $v$ values, $C_{11}$ is an increasing function of $v$. $C_{11}$ has a broad maximum around $v=0.44$, and a broad minimum around $v=0.81$. $C_{11}$ becomes smaller as $d/l_0$ is increasing even for higher filling factors. The shear moduli as a function of $v$ for $d/l_0 = 0$ and 0.7 are shown in figures 6 and 7, respectively. $C_{66}$ has a broad maximum around $v = v_{max}$, and turns decreasing for $v > v_{max}$. Values of $v_{max}$ for $d/l_0 = 0$ and 0.7 are $v=0.33$ and 0.31, respectively. Therefore, $v_{max}$ has the tendency toward smaller filling factors if the sample width becomes higher. Above a transition filling factor $v_0$, $C_{66}$ becomes negative, implying that the dipolar-particle crystal is unstable for $v > v_0$ and stable in the regime $0 < v \leq v_0$. Since the shear modulus of the hole crystal is symmetrical to that of the particle crystal, the hole crystal is locally stable for $1 - v_0 \leq v < 1$. Similar results were discovered by Maki and Zotos in the electronic crystal [31]. We obtain filling factors of $\nu_0 = 0.497$ and 0.478 for $d/l_0 = 0$ and 0.7, respectively.

From the elastic moduli, we can derive collective modes of the WC. We shall decompose the position vector $\mathbf{r}_i$ of the *i*th particle into $\mathbf{r}_i = \mathbf{R}_i + \mathbf{u}_i$, where $\mathbf{u}_i$ is the displacement vector of the *i*th particle from the equilibrium lattice site $\mathbf{R}_i$. In the harmonic approximation the dynamical equations for displacements $u_{i\alpha}$ along the $\alpha$-axis can be written in the form

$$M \frac{d^2}{dt^2} u_{i\alpha} = -\sum_{j\beta} \Phi_{\alpha\beta} (\mathbf{R}_i - \mathbf{R}_j) u_{j\beta} + M\omega_c \varepsilon_{\alpha\beta} \frac{d}{dt} u_{i\alpha}, \qquad (32)$$

where $\varepsilon_{\alpha\beta}$ is the antisymmetric tensor in two dimensions and the last term in this equation corresponds to the Coriolis force from rotation. We shall find a solution to the above dynamical equation by taking $u_{i\alpha} = A_{i\alpha}(\mathbf{k}) \exp[i(\mathbf{k} \cdot \mathbf{R}_i - \varpi t)]$ that represents a wave with amplitude $A_{i\alpha}(\mathbf{k})$, angular frequency $\varpi$ and wave vector $\mathbf{k}$. Substituting the above expression into Eq. (32) results in the following secular equation of the WC:

$$\begin{pmatrix} \tilde{\Phi}_{xx}(\mathbf{k}) - \varpi^2 & \tilde{\Phi}_{xy}(\mathbf{k}) - i\varpi\omega_c \\ \tilde{\Phi}_{yx}(\mathbf{k}) + i\varpi\omega_c & \tilde{\Phi}_{yy}(\mathbf{k}) - \varpi^2 \end{pmatrix} \begin{pmatrix} A_x \\ A_y \end{pmatrix} = 0, \qquad (33)$$

where $\tilde{\Phi}_{\alpha\beta}(\mathbf{k}) = \Phi_{\alpha\beta}(\mathbf{k})/M$.

To find eigenfrequencies of collective modes we have to solve the vanishing determinant of the above secular equation. From Eq. (33) we obtain the eigenfrequency of phonons

$$\varpi_\pm^2 = \frac{\mathrm{tr}[D(\mathbf{k})] + \omega_c^2}{2} \pm \frac{1}{2}\sqrt{\left(\mathrm{tr}[D(\mathbf{k})] + \omega_c^2\right)^2 - 4\det[D(\mathbf{k})]}, \qquad (34)$$

where $\text{tr}[D(\mathbf{k})] = \tilde{\Phi}_{xx}(\mathbf{k}) + \tilde{\Phi}_{yy}(\mathbf{k})$ and $\det[D(\mathbf{k})] = \tilde{\Phi}_{xx}(\mathbf{k})\tilde{\Phi}_{yy}(\mathbf{k}) - \tilde{\Phi}_{xy}(\mathbf{k})\tilde{\Phi}_{yx}(\mathbf{k})$. In the absence of the Coriolis force, one finds, in the long-wavelength limit and wave vector $k \to 0$, a longitudinal phonon mode with frequency $\omega_\ell \sim \sqrt{C_{11}}\, k\omega_d$ and a transverse phonon mode with frequency $\omega_t \sim \sqrt{C_{66}}\, k\omega_d$, where $\omega_d = \sqrt{D/M\ell^5}$. The linear dispersion of the longitudinal phonon mode is due to the nature of the dipolar force and survives in the liquid phase, while the transverse phonon mode is specific to the crystalline phase. These two modes are coupled by the presence of the Coriolis force on the vibrating particles. In the limit $\omega_c \gg \omega_\ell$ and $\omega_c \gg \omega_t$, two coupled modes give rise to a low-frequency $\omega_- \sim \omega_\ell \omega_t / \omega_c$ and to a high-frequency mode $\omega_+ \sim \omega_c$. At long wavelength the low-frequency mode has a quadratic dispersion relation, $\omega_- \sim \sqrt{C_{11}C_{66}}\, k^2 \omega_d^2 / \omega_c$.

## V. QUANTUM FLUCTUATIONS OF THE WIGNER CRYSTAL

In this section, we use the harmonic approximation to determine the mean square of the displacement vector at absolute zero temperature. The average displacement is given by use of the dynamical matrix.

We will apply the Lindeman's criterion to examine the stability of the crystal, i.e. the crystal will melt if the mean square of the displacement vector $u$ exceeds some critical value. The average displacement $\Gamma$ per unit of the lattice constant $a$ is defined by

$$\Gamma = \sqrt{\langle u^2 \rangle / a^2}. \tag{35}$$

We consider the crystal being unstable if $\Gamma > 0.15$. In the following we will evaluate $\Gamma$ in the harmonic approximation. From Eq. (34), the mean square of the displacement vector at zero temperature [35] is given by

$$\left\langle u^2 \right\rangle = \frac{\hbar}{2MN} \sum_{\mathbf{k}} \frac{(\omega_\ell + \omega_t)^2}{\omega_+ \omega_- (\omega_+ + \omega_-)}, \tag{36}$$

where $N$ is the particle number.

In the explicit evaluation we will employ the Debye model with the cut-off momentum $k_D = \sqrt{4\pi\rho}$.

In the approximation $\omega_c \gg \omega_D$, where $\omega_D$ is the Debye frequency of the crystal, we obtain the mean square of the displacement of a particle crystal expressed in terms of elastic moduli:

$$\frac{\left\langle u^2 \right\rangle}{a^2} = \frac{\sqrt{3}\nu}{8\pi} \frac{\left(\sqrt{C_{11}} + \sqrt{C_{66}}\right)^2}{\sqrt{C_{11}C_{66}}}. \tag{37}$$

The mean square of the displacement of a hole crystal can be also obtained if the value $\nu$ in Eq. (37) is replaced by the value $1-\nu$. Equation (37) implies that quantum fluctuations of particle and hole crystals are enhanced in the region of large $\nu$ and $1-\nu$, respectively. This enhancement of quantum fluctuations will reduce the region where the crystal is stable. In figures 8 and 9 we show the average displacement $\Gamma$ as a function of $\nu$ from a particle crystal. The WC will melt if $\Gamma > 0.15$. The critical value that the WC will start melting is at the transition point $\nu = 0.66$ ($\sim \frac{1}{15}$), which is the sample thickness independent. This result indicates that rotating dipolar fermions with $\nu < \frac{1}{15}$ will crystallize into a particle crystal. As we have shown that the hole crystal has the lowest correlation energy for $\nu > \frac{1}{2}$, we can also conclude that rotating dipolar fermions with $\frac{14}{15} < \nu < 1$ will crystallize into a hole crystal. Such possibility of crystallization has been discussed by Baranov, Fehrmann and Lewenstein [27]. They conclude that rotating dipolar fermions will crystallize into a WC in a region as $\nu < \frac{1}{7}$, where the stable region of the WC is larger than our study.

## VI. CONCLUSIONS

In conclusion, in this paper we have calculate the correlation energies, compression and shear moduli, and collective modes of the WC in the lowest Landau level. The effect of finite sample thickness has been included in studying the physical properties of the WC. We find that this effect reduces the correlation energies and elastic moduli of the WC. Going beyond previous treatments and by the Löwdin formula, we can calculate accurate correlation energies of the WC in the whole lowest Landau-level regime. We find that the correlation energy of a particle crystal is lower and higher than the correlation energy of a hole crystal for $\nu < \frac{1}{2}$ and $\nu > \frac{1}{2}$, respectively. We have also examined the stability of the WC from the Lindeman's criterion that the crystal will melt if the mean square of the displacement vector exceeds some critical value. In a region $\nu < \frac{1}{15}$, the mean displacement vector of the WC is less than 0.15, implying that the WC is stable. The particle and hole crystals are stable in regions $0 < \nu \leq \frac{1}{15}$ and $\frac{14}{15} \leq \nu < 1$, respectively, where the precise value of the transition filling factor is independent on the sample thickness.

### ACKNOWLEGEMENTS

We acknowledge the partial financial support from the National Science Council (NSC) of Republic of China under Contract No. NSC96-2112-M-034-002-MY3. The author also thanks the support of the National Center for Theoretical Sciences of Taiwan during visiting the center.

**REREFENCES**

[1] F. Dalfovo, S. Giorgini, L. P. Pitaevskii, and S. Stringari, Rev. Mod. Phys. **71**, 463 (1999).

[2] A. S. Parkins and D. F. Walls, Phys. Rep. **303**, 1 (1998).

[3] A. J. Leggett, Rev. Mod. Phys. **73**, 307 (2001).

[4] A Griesmaier, J. Werner, S. Hensler, J. Stuhler and T. Pfau, Phys. Rev. Lett. **94**, 160401 (2005).

[5] J. Stuhler, A. Griesmaier, T. Koch, M. Fattori, T. Pfau, S. Giovanazzi, P. Pedri and L. Santos, Phys. Rev. Lett. **95**, 150406 (2005).

[6] A. J. Kerman, J. M. Sage, S. Sainis, T. Bergeman and D. DeMille, Phys. Rev. Lett. **92**, 033004 (2004).

[7] C. A. Stan, M. W. Zwierlein, C. H. Schunck, S. M. F. Raupach and W. Ketterle, Phys. Rev. Lett. **93**, 143001 (2004).

[8] S. Inouye, J. Goldwin, M. L. Olsen, C. Ticknor, J. L. Bohn and D. S. Jin, Phys. Rev. Lett. **93**, 183201 (2004).

[9] J. M. Sage, S. Sainis, T. Bergeman, and D. DeMille, Phys. Rev. Lett. **94**, 203001 (2005).

[10] C. Ospelkaus, S. Ospelkaus, L. Humbert, P. Ernst, K. Sengstock and K. Bongs, Phys. Rev. Lett. **97**, 120402 (2006).

[11] J. Kleinert, C. Haimberger, P. J. Zabawa and N. P. Bigelow, Phys. Rev. Lett. **99**, 143002 (2007).

[12] K. Góral, K. Rzazewski, and T. Pfau, Phys. Rev. A **61**, 051601(R) (2000).

[13] K. Osterloh, N. Barberan, and M. Lewenstein, Phys. Rev. Lett. **99**, 160403 (2007).

[14] H. P. Büchler, E. Demler, M. Lukin, A. Micheli, N. Prokof'ev, G. Pupillo and P. Zoller, Phys. Rev. Lett. **98**, 060404 (2007).

[15] G. E. Astrakharchik, J. Boronat, I. L. Kurbakov, and Y. E. Lozovik, Phys. Rev. Lett. **98**, 060405 (2007).

[16] U. R. Fischer, Phys. Rev. A **73**, 031602(R) (2006).

[17] N. R. Cooper, E. H. Rezayi, and S. H. Simon, Phys. Rev. Lett. **95**, 200402 (2005).

[18] J. M. V. A. Koelman, H. T. C. Stoof, B. J. Verhaar, and J. T. M. Walraven, Phys. Rev. Lett. **59**, 676 (1987).

[19] K. M. O'Hara, S. L. Hemmer, M. E. Gehm, S. R. Granade and J. E. Thomas, Science **298**, 2179 (2002).

[20] T. Bourdel, J. Cubizolles, L. Khaykovich, K. M. F. Magalhães, S. J. J. M. F. Kokkelmans, G. V. Shlyapnikov and C. Salomon, Phys. Rev. Lett. **91**, 020402 (2003).

[21] C. A. Regal, M. Greiner, and D. S. Jin, Phys. Rev. Lett. **92**, 040403 (2004).

[22] M. W. Zwierlein, C. A. Stan, C. H. Schunck, S. M. F. Raupach, A. J. Kerman and W. Ketterle, Phys. Rev. Lett. **92**, 120403 (2004).

[23] S. Giorgini, L. P. Pitaevskii, and S. Stringari, Rev. Mod. Phys. **80**, 1215 (2008).

[24] M. A. Baranov, Phys. Rep. **464**, 71 (20088).

[25] M. A. Baranov, L. Dobrek, and M. Lewenstein, Phys. Rev. Lett. **92**, 250403 (2004).

[26] G. M. Bruun, and E. Taylor, Phys. Rev. Lett. **101**, 245301 (2008).

[27] M. A. Baranov, H. Fehrmann, and M. Lewenstein, Phys. Rev. Lett. **100**, 200402 (2008).

[28] E. Wigner, Phys. Rev. **46**, 1002 (1934).

[29] L. D. Landau and E. M. Lifshitz, *Quantum Mechanics*, Pergamon Press, Oxford, New York, (1977).

[30] K. v.Klitzing, G. Dorda, and M. Pepper, Phys. Rev. Lett. **45**, 494 (1980).

[31] K. Maki and X. Zotos, Phys. Rev. B **28**, 4349 (1983).

[32] S. C. Cheng, Chinese Journal of Physics **35**, 718 (1997).

[33] L. Bonsall and A. A. Maradudin, Phys. Rev. B **15**, 1959 (1977).

[34] P. O. Löwdin, J. Chem. Phys. **18**, 365 (1950).

[35] H. Fukuyama, Solid State Commun. **19**. 551 (1976).

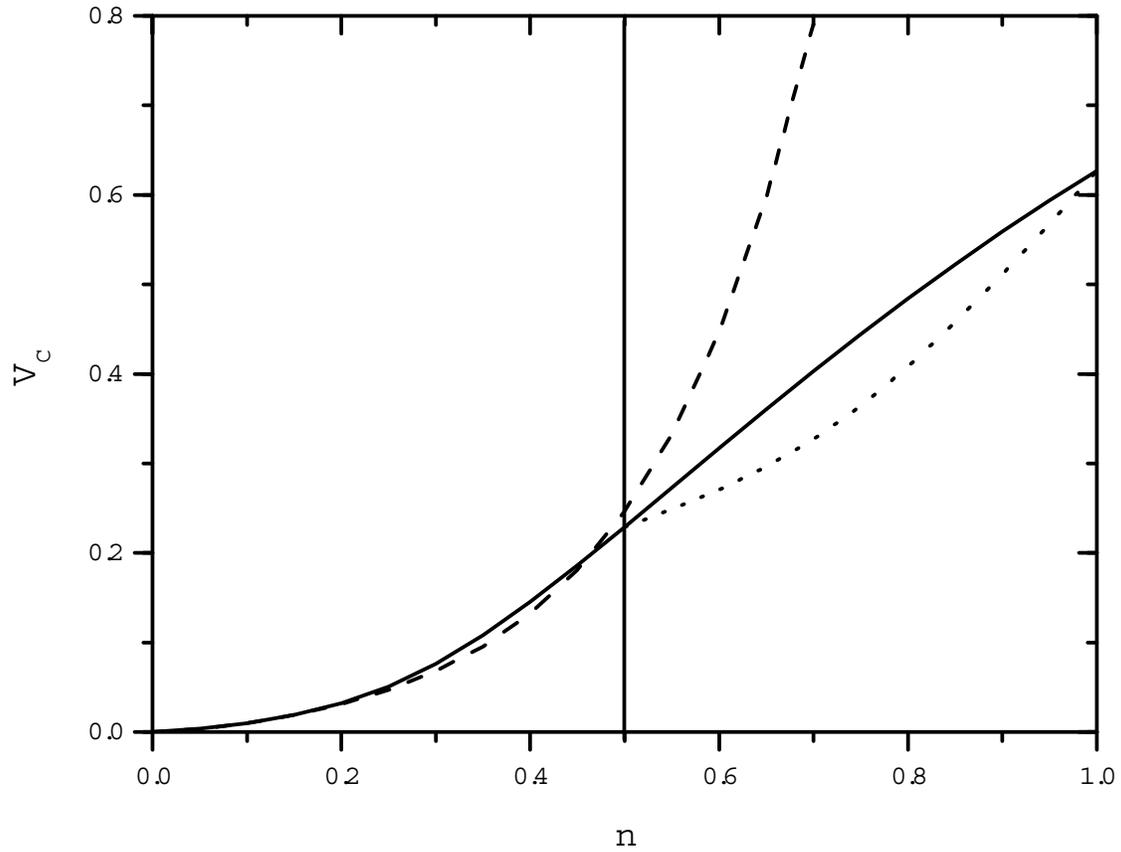

Fig. 1. Correlation energy per particle in units of $D/l_0^3$ is shown as a function of the filling factor. The solid line is for the particle crystal, the dotted line is for the hole crystal, and the dashed line is taken from Ref. 27. The sample thickness $d/l_0 = 0$.

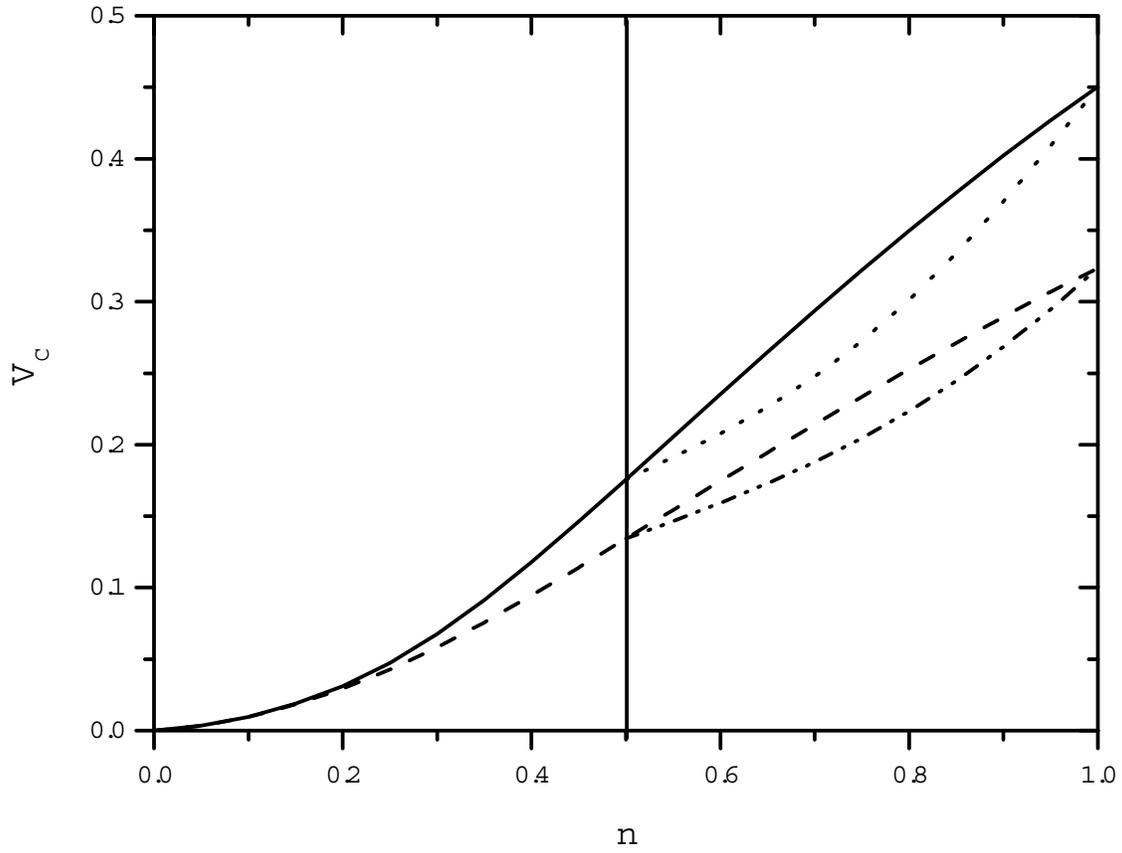

Fig. 2. Correlation energy per particle in units of $D/l_0^3$ is shown as a function of the filling factor. The solid (dotted) line is for the particle (hole) crystal with thickness $d/l_0 = 0.3$. The dashed (dashed-dotted) line is for the particle (hole) crystal with thickness $d/l_0 = 0.7$.

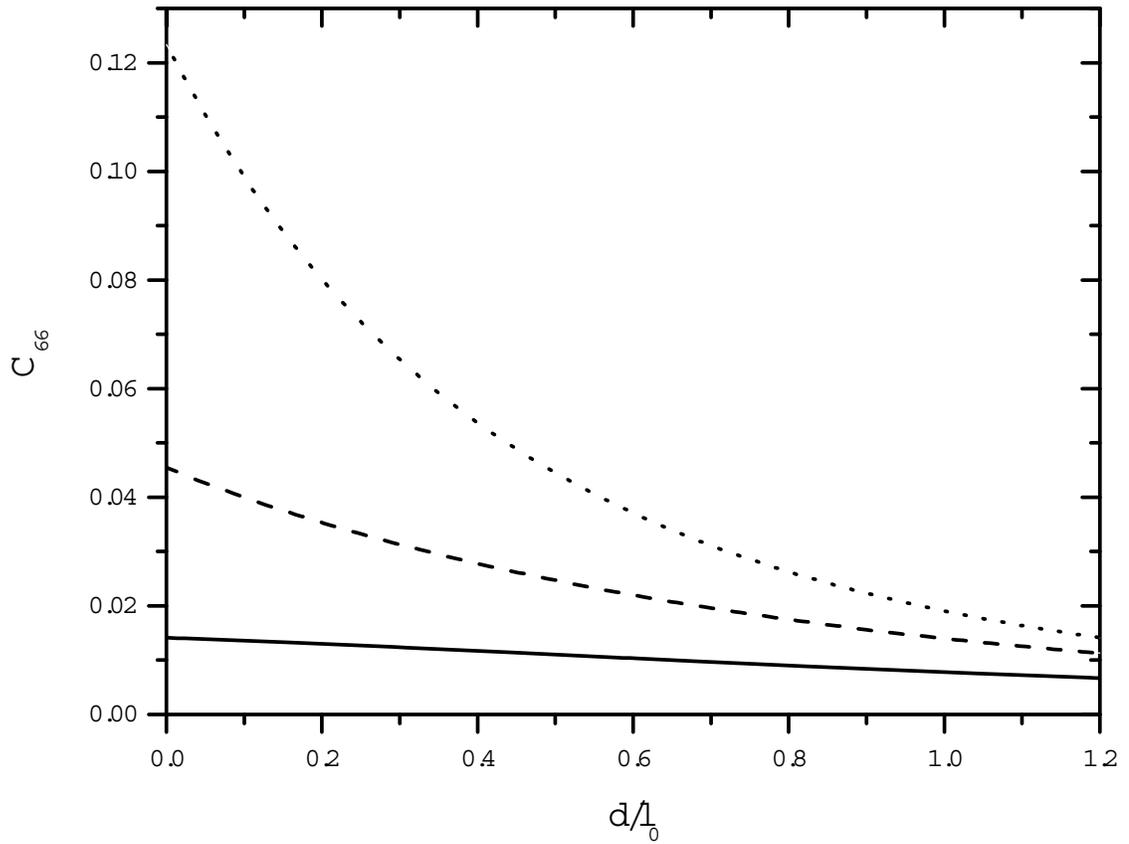

Fig. 3. Shear moduli of Wigner crystals in units of $D/l_0^3$ is shown as a function of the sample thickness. The solid, dashed, and dotted lines are for particle crystals with filling factors $\nu=1/7$, 1/5 and 1/3, respectively.

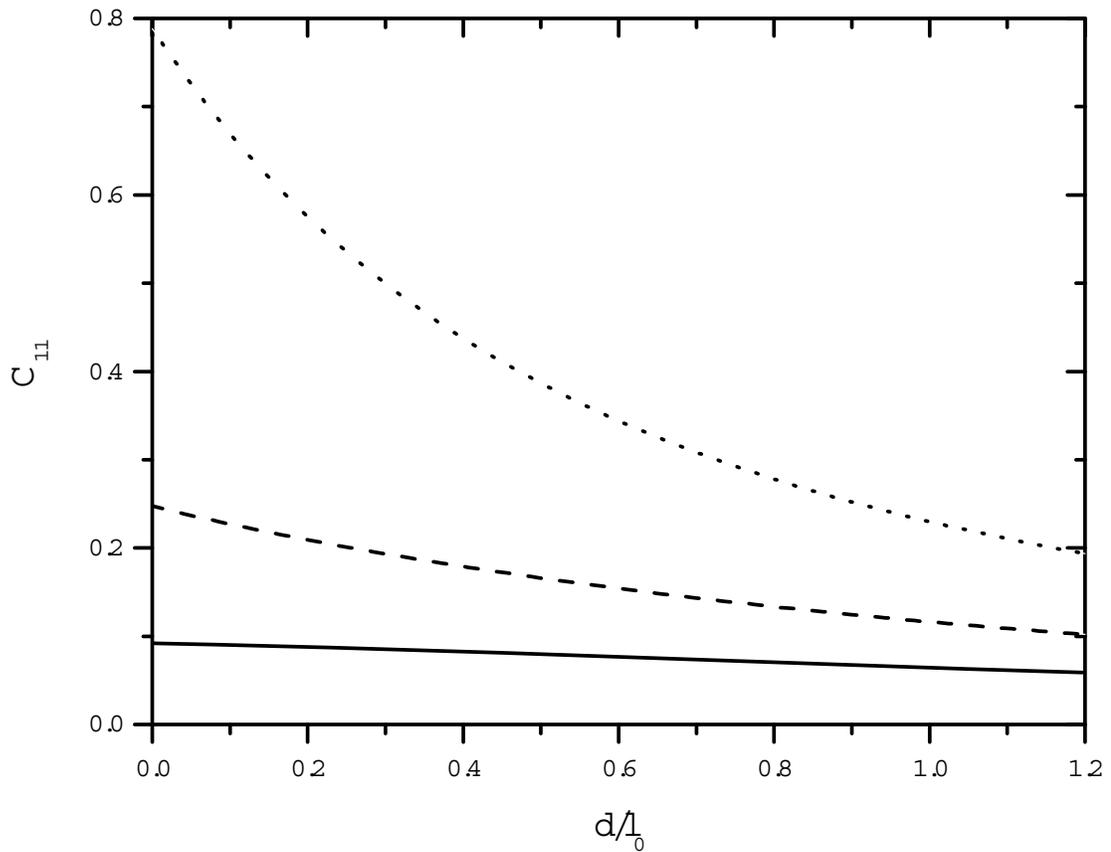

Fig. 4. Compression moduli of Wigner crystals in units of $D/l_0^3$ are shown as function of the sample thickness. The solid, dashed, and dotted lines are for particle crystals with filling factors $\nu=1/7$, 1/5, and 1/3, respectively.

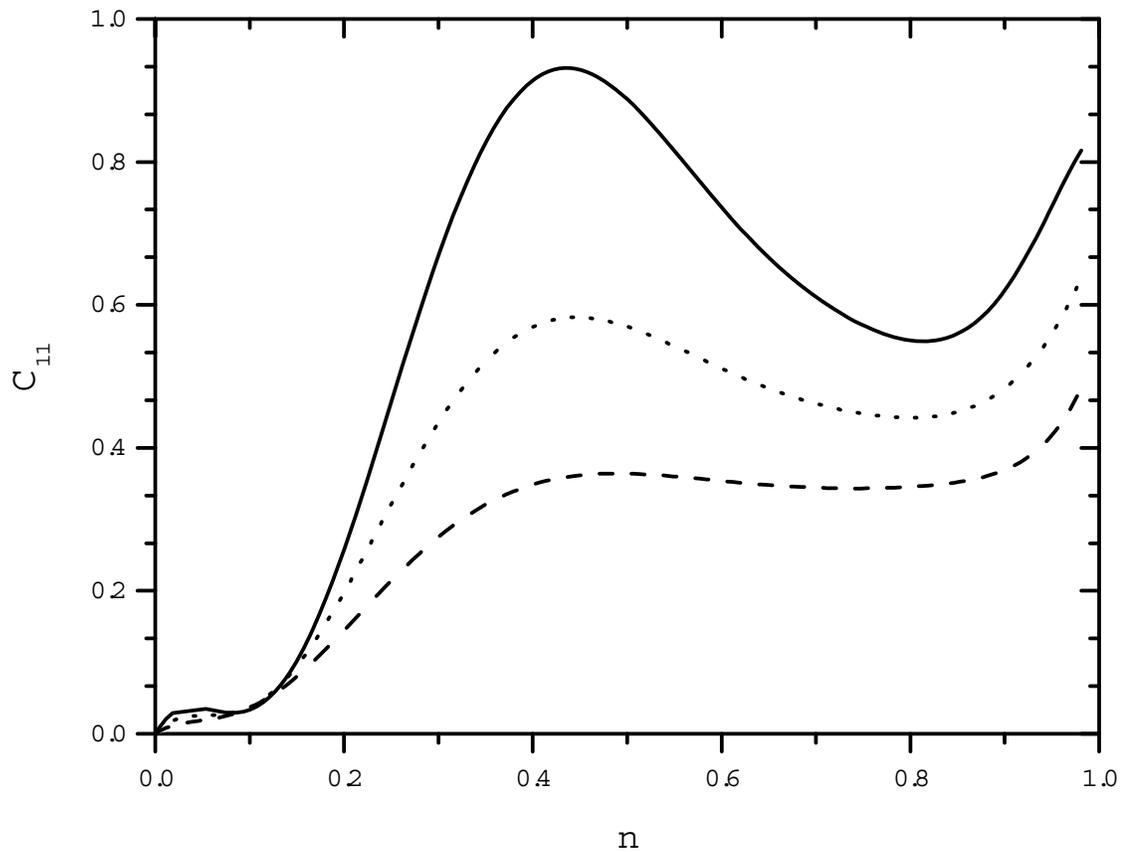

Fig. 5. Compression moduli of Wigner crystals in units of $D/l_0^3$ are shown as function of the filling factor. The solid, dotted, and dashed lines are for particle crystals with thicknesses $d/l_0 = 0$, 0.3, and 0.7, respectively.

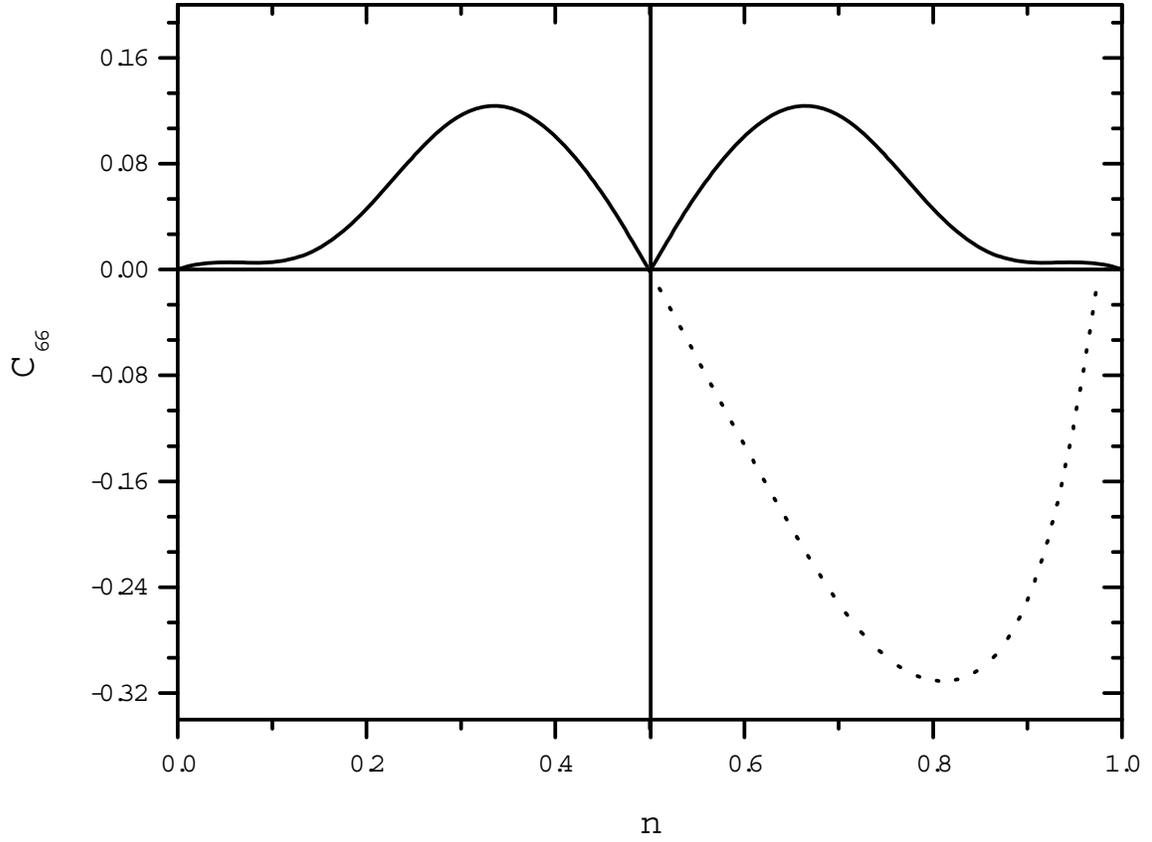

Fig. 6. Shear moduli, in units of $D/l_0^3$, of Wigner crystals with thickness $d/l_0 =0$ are shown as a function of the filling factor. The solid line for $\nu<0.5$ and the dotted line are the shear moduli of a particle crystal. The solid line for $\nu>0.5$ is the shear moduli of a hole crystal.

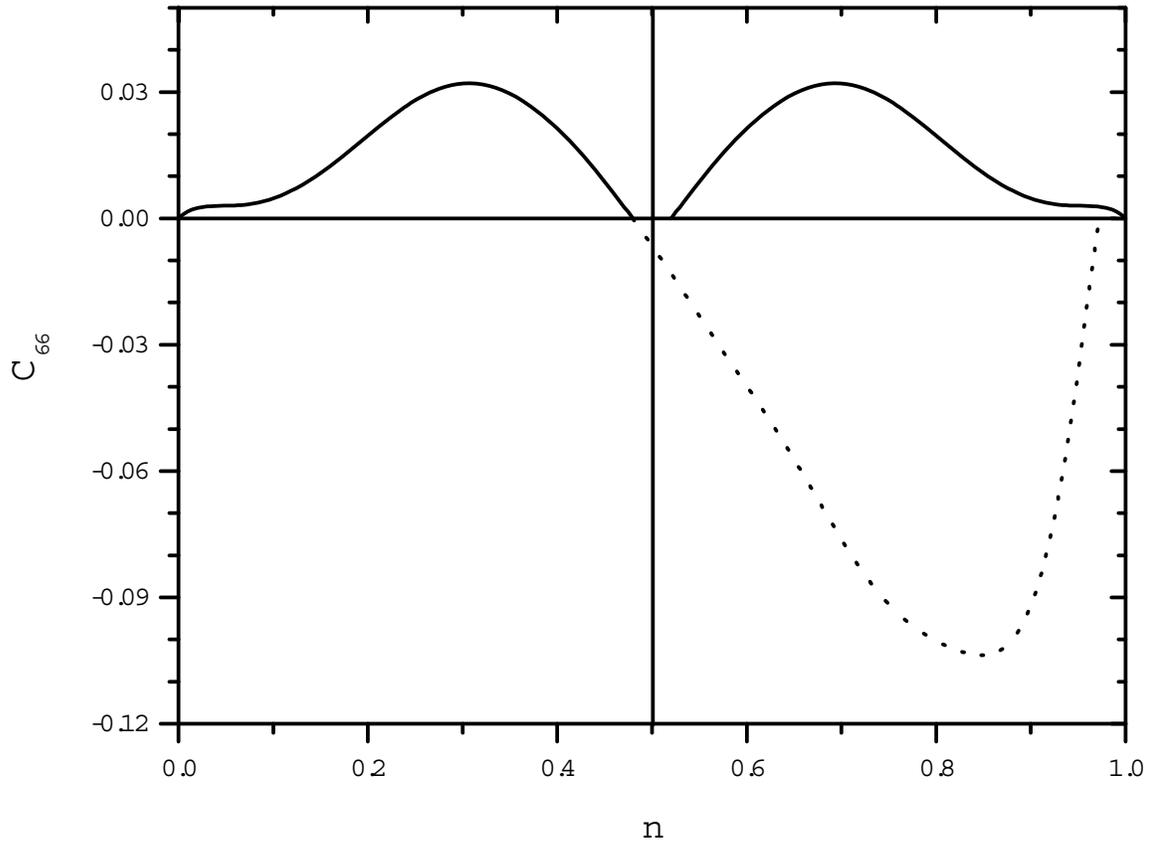

Fig. 7. Shear moduli, in units of $D/l_0^3$, of Wigner crystals with thickness $d/l_0 = 0.7$ are shown as a function of the filling factor. The solid line for $\nu < 0.5$ and the dotted line are the shear moduli of a particle crystal. The solid line for $\nu > 0.5$ is the shear moduli of a hole crystal.

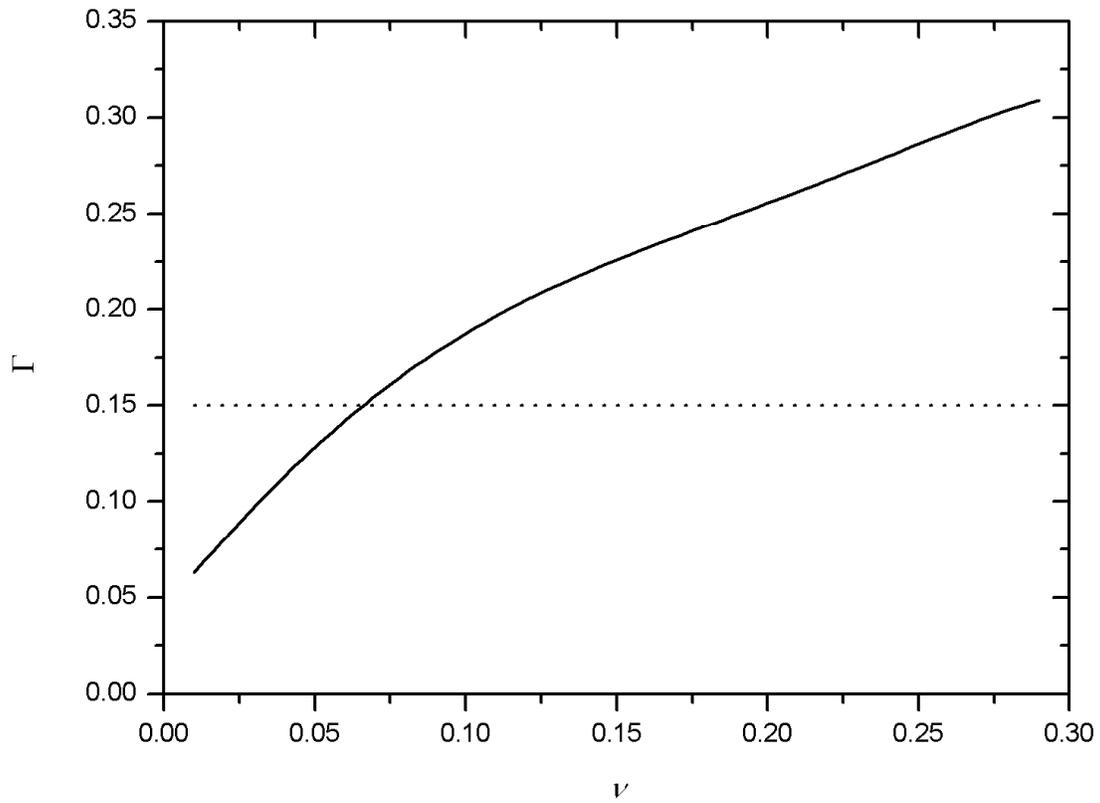

Fig. 8. Mean displacement, in units of *a*, of Wigner crystals with thickness $d/l_0 = 0$ is shown as a function of the filling factor. The solid line is the average displacement of a particle crystal. The dotted line indicates the critical value of melting.

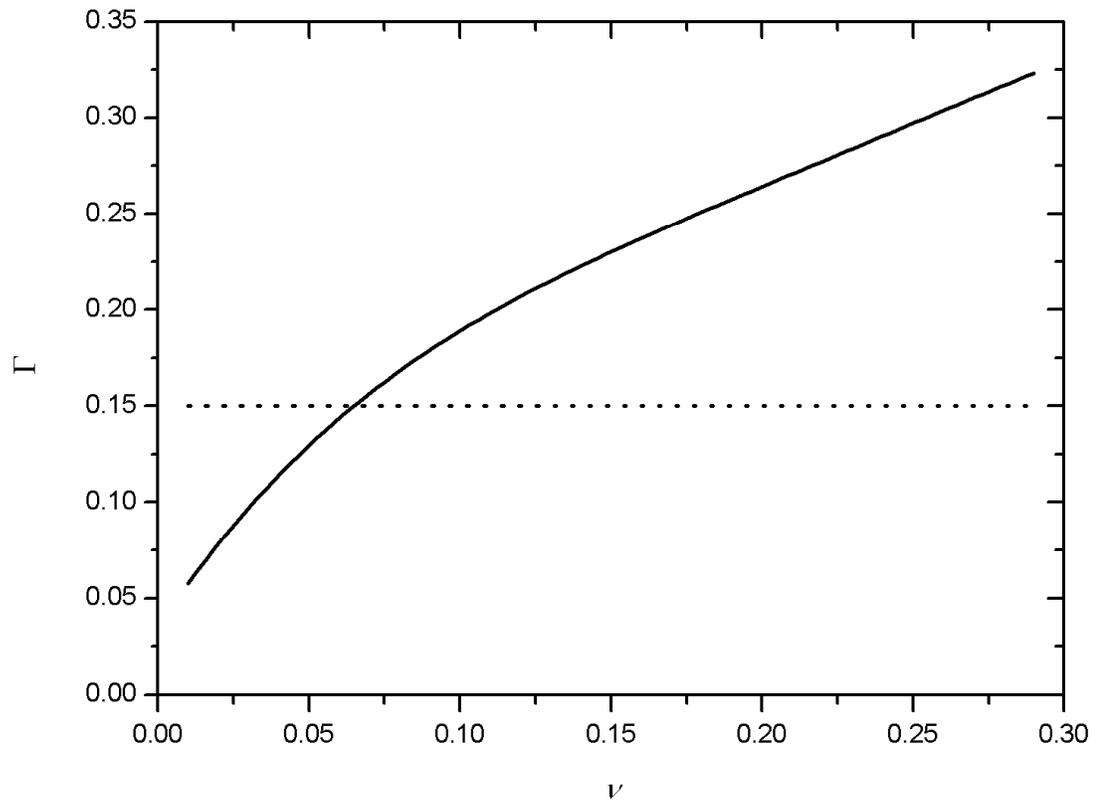

Fig. 9. Mean displacement, in units of *a*, of Wigner crystals with thickness $d/l_0$ =0.7 is shown as a function of the filling factor. The solid line is the average displacement of a particle crystal. The dotted line indicates the critical value of melting.